\documentclass[prl,aps,twocolumn,epsfig]{revtex4}
\usepackage{epsfig}
\usepackage{wasysym}
\usepackage{bm}
\newlength{\myL}

\newcommand{\beq}{\begin{equation}}
\newcommand{\eeq}{\end{equation}}
\newcommand{\bea}{\begin{eqnarray}}
\newcommand{\eea}{\end{eqnarray}}

\newcommand{\bB}{{\bf{B}}}
\newcommand{\bA}{{\bf{A}}}

\newcommand{\wernercomment}[1]{}

\def\subsub#1{\noindent{\bf #1:}}

\def\tit#1#2#3#4#5{{#1}{\bf #2}, #3 (#4)}
\def\jmp{J.\ Math.\ Phys.\ }

\def\rmp{Rev.\ Mod.\ Phys.\ }

\def\prl{Phys.\ Rev.\ Lett.\ }
\def\pr{Phys.\ Rev.\ }
\def\prb{Phys.\ Rev.\ B\ }

\def\jpa{J.\ Phys.\ A\ }

\def\jpsj{J.\ Phys.\ Soc.\ Jpn.\ }

\def\jsp{J.\ Stat.\ Phys.\ }

\def \x {{\bf x}}

\begin{document}


\title{Coulomb and Liquid Dimer Models in Three Dimensions}

\author{David A. Huse,$^1$ Werner Krauth,$^2$ 
R. Moessner$^3$ and S. L. Sondhi$^1$}

\affiliation{$^1$Department of Physics, Princeton University,
Princeton, NJ 08544, USA}

\affiliation{$^2$CNRS-Laboratoire de Physique Statistique de l'Ecole Normale
Sup\'erieure, Paris, France}

\affiliation{$^3$Laboratoire de Physique Th\'eorique de l'Ecole Normale
Sup\'erieure, CNRS-UMR8549, Paris, France}

\date{\today}

\begin{abstract}
We study classical hard-core dimer models on three-dimensional lattices
using analytical approaches and Monte Carlo simulations.  On the
bipartite cubic lattice, a local gauge field generalization of the height
representation used on the square lattice predicts that the dimers are
in a critical Coulomb phase with algebraic, dipolar, correlations, in
excellent agreement with our large-scale Monte Carlo simulations. The
non-bipartite FCC and Fisher lattices lack such a representation, and we
find that these models have both confined and exponentially deconfined
but no critical phases.  We conjecture that extended critical phases
are realized only on bipartite lattices, even in higher dimensions.
\end{abstract}


\maketitle


The statistical mechanics of dimers on a lattice that interact with one
another only via hard-core exclusion has long been of interest
to mathematicians and physicists \cite{kenyonrev,dombdim}. It is one of
the simplest models describing the arrangement of anisotropic objects on
a regular substrate.  Applications \cite{fywu2003} range from diatomic
molecules on surfaces to spin ice in a magnetic field \cite{ogata}.

By Kasteleyn's theorem, on two-dimensional ($2d$) planar lattices, the statistical
mechanics of (close-packed) dimer coverings can be computed
exactly \cite{kasteleyn}.  A consistent picture has emerged from this
work for a large class of $2d$ dimer models.  On bipartite $2d$ lattices,
dimer models are in confined phases in which the free energy of two
inserted test monomers (unpaired sites) increases with separation. The
increase is logarithmic for phases with algebraic dimer correlations
and linear for the remaining ones.  An example of the former is the
square lattice \cite{kasteleyn,fisher63} and one of the latter is the
exotic diamond-octagon ``4-8'' lattice which exhibits two confining
phases as the strength of the diamond bonds is varied \cite{fywu2003}.  By contrast,
non-bipartite lattices exhibit both deconfined and confined phases but always with
exponentially decaying dimer correlations except at the boundary between
such phases. Examples are the triangular \cite{fendley,mcdimer} and 
kagome \cite{gregkag} lattices, which are deconfined
with exponentially decaying correlations, and the Fisher lattice --
equivalent to the $2d$ Ising model \cite{fishlat} -- which exhibits a
deconfinement transition \cite{fishdim}.

Dimer models on two-dimensional bipartite lattices can also be understood
through their height representations \cite{Blote82,henleyjsp}.
Within this powerful framework,  the two sub-categories of critical
and non-critical dimer correlations are described as  ``rough" and
``flat" phases.  In either case the defect interaction corresponding to
the monomer free energy is long-ranged.  Dimer models on non-bipartite
lattices lack a similar long-wavelength description.

In this paper we generalize the height representation to three dimensions.
We show that dimer models on bipartite $3d$ lattices admit a local
gauge representation which results in a ``Coulomb'' phase with algebraic,
dipolar forms for the dimer correlations and monomer interactions that
fall off inversely with their separation. We present large scale Monte
Carlo simulations on $3d$ lattices and demonstrate that the dimer
model on the cubic lattice is critical.  On non-bipartite lattices
we show that  there exist  both confined and deconfined phases, with
exponentially decaying dimer correlations and monomer interactions,
finding that the face-centered cubic (FCC) lattice and the $3d$
Fisher lattice realize these cases. As will be discussed
elsewhere, the Coulomb and liquid phases can be identified with the
$U(1)$ and $Z_2$ deconfined phases \cite{wip} in the corresponding
quantum dimer models \cite{Rokhsar88}.

We first report our results on the cubic lattice along with the gauge
representation. Thereafter we consider the FCC and Fisher lattices and
conclude with a summary and some conjectures on related models.

\subsub{Cubic lattice}
Generally, for any dimer configuration, we  define dimer numbers ${\bf n}
(\x)$ as $n_i(\x)=1$ if the bond between sites $\x$ and $\x + {\bf e}_i$
is occupied, and zero otherwise  (${\bf e}_i$ is the unit vector in
direction $i$).  Close-packed hard-core dimers obey the condition
$\sum_i n_i (\x) + n_i (\x - {\bf e}_i) = 1$.

\begin{figure}[htbp]
\begin{center}
\epsfig{file=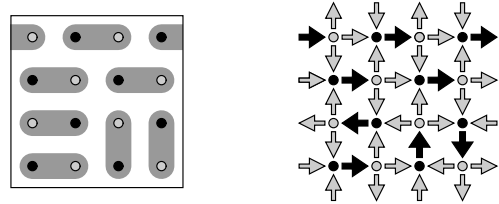,width=0.75\columnwidth}
\end{center}
\caption{
Dimer configuration on the bipartite square lattice (left)
and corresponding configuration of the divergence-free magnetic field 
(right---a dark arrow carries $3$ times the flux of a light one).
The two lower
quadrants represent plaquettes which can be flipped 
(dimers rotated by $ \pi/2$).
These plaquettes have zero average $\bB$-field.  }
\label{fig:dimer_conf}
\end{figure}

On a bipartite lattice, each dimer touches one site on each  sublattice
(cf Fig.~\ref{fig:dimer_conf}, where the two sublattices are indicated
by light and dark dots).  Using a sublattice sign factor $\sigma_\x =
\pm 1$ depending on whether the site $\x$ belongs to one or the other
sublattice, we can now define a field variable on each link
\beq
B_i (\x) = \sigma_{\x} ( n_i (\x) - z^{-1})\ 
\eeq
(where $z$ is the coordination number) as shown 
in  Fig.~\ref{fig:dimer_conf}.  
For close-packed dimers the lattice divergence of the field $\bB$
vanishes, $\sum_i B_i (\x) - B_i (\x - {\bf e}_i) = 0$ i.e., $\bB$
is a lattice magnetic field without monopoles. A monomer is a monopole
with charge $\pm 1$ depending on the sublattice.

We  define the lattice flux $\phi_\Sigma$ through a surface $\Sigma$
which does not contain any sites as  the sum of the magnetic fields on the
links piercing it $\phi_\Sigma=( \iint \bB \cdot {\bf dS})_{\rm lattice}$.
For a cube with periodic boundary conditions, the flux through any surface
that wraps around the system is invariant under local rearrangements of
the dimers and under lattice translations of the surface.  In particular,
if we let $\Sigma_i$ be planes perpendicular to the cubic unit vectors
${\bf e}_i$, the fluxes $\Phi_i$ through them are the maximal invariants
that characterize a given topological sector of the dimer model.  For an
$L^3$ cube the maximal possible flux is $L^2/2$.

A dimer configuration may  be represented by a lattice
magnetic field in any dimension.  In $2d$, one solves the constraint
$\nabla \cdot \bB =0$ through $\bB = \nabla \times h$ \cite{youngaxe}
where $h$, the height function mentioned earlier, is a scalar field on
the dual lattice. In $3d$, $\bB$ is computed from a vector
potential $\bA$ on the links of the dual lattice. Given an arbitrary
$\bA$ we recover $\bB$ as its (lattice) curl, $\bB = \nabla \times
\bA$ which is computed as \beq B_i(\x) = \sum_{{\bf y} \epsilon
\partial p} A_j(\bf y), \eeq the oriented sum of the link variables on
the boundary $\partial p$ of the dual plaquette $p_i(\x)$ pierced by the
link $(\x, \x+ {\bf e}_i)$.  While in $2d$,  $h$ is defined up
to a global constant, in $3d$,  we have local gauge
transformations, $A_j({\bf y}) \rightarrow A_j({\bf y}) + \Lambda({\bf
y} + {\bf e}_j) - \Lambda({\bf y}) $, 
where $\Lambda$ is an arbitrary function on the sites of
the dual lattice. Consequently, it is necessary to pick a gauge to
work out properties of the $\bA$. 
The fluxes $\Phi_i$ can  be computed from the lattice line
integrals of the vector potential along the boundaries of
$\Sigma_i$. Hence the sector with all $\Phi_i =0$ is obtained from
gauge fields $\bA$ that obey periodic boundary conditions themselves.

In analogy to $2d$ we now conjecture that the long wavelength fluctuations
of $\bA$ and therefore $\bB$ are governed by a probability distribution
for the coarse-grained fields:
\bea
P[\bA] \propto e^{-{K \over 2} \int_V (\nabla \times \bA)^2}
\equiv e^{-{K \over 2} \int_V \bB^2}
\label{e:conj}
\eea
in the vicinity of the zero flux $\Phi_i =0$ state. In the exponent
of Eq.~(\ref{e:conj}) the energy $\bB^2/2$ of a magnetic field appears
naturally.  Configurations that locally minimize the (coarse-grained) field
strength (cf  the lower quadrants of Fig.~\ref{fig:dimer_conf}) maximize
the number of flippable plaquettes with two parallel dimers and have
high entropy, as described by Eq.~{\ref{e:conj}}.

\noindent Two comments are in order: \\
\emph{i)} The assertion Eq.~\ref{e:conj} implies that the gauge field
is in a Coulomb phase, in the language of lattice gauge theories. The
existence of this phase in our lattice system is not in conflict with
Polyakov's proof of confinement \cite{polymono} 
for the standard U(1) lattice gauge
theory in $3d$ because in our case the microscopics explicitly forbid
the monopoles that were crucial to his analysis.\\
\emph{ii)} Gauge invariance explictly forbids any relevant operators
at the fixed point defined by Eq.~\ref{e:conj}; this is the standard
explanation of the masslessness of the photon. Consequently the prediction
of a Coulomb phase is self-consistent and weak perturbations cannot give
rise to anything new. This should be contrasted with the situation in
$2d$, where vertex operators can become relevant even at weak coupling
and, depending on the height stiffness and the radius of the height
field, lead to a flat phase instead of the rough phase described by a
purely Gaussian action.

Returning to the ansatz (\ref{e:conj}) it is straightforward to 
deduce the long distance correlator,
\beq
\langle B_i(\x) B_j(0) \rangle = \frac{1}{4\pi K} {3 x_i x_j - r^2 \delta_{ij}
\over r^5}
\label{e:dipole}
\eeq
which is of the standard $3d$ dipole form. Reinserting the
sublattice sign factors gives the connected dimer correlators.

\begin{figure}[htbp]
\begin{center}

\epsfig{file=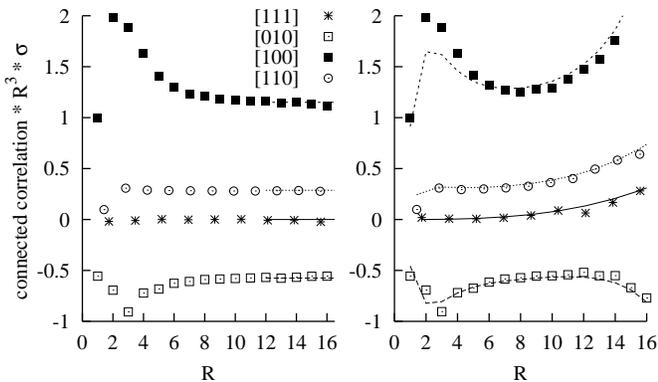,width=\columnwidth}
\end{center}
\caption{Monte Carlo data of connected correlations (symbols) between
parallel dimers, $z^2 \langle n_i(\x)n_i(0)\rangle -1$, in various
directions, multiplied by the sublattice sign factor $\sigma_{\x}$ and
by $R^3$, plotted \emph{vs.\ }the Euclidean distance between the
dimers, $R$. The coordination $z=6$ for the cubic lattice.  Left: For
a system of size $L=128, R \ll L $, the dipolar asymptotes are
indicated by horizontal lines. The overall scale factor is fixed by
normalising those lines with respect to the $[100]$ correlations.
Right: For a system size $L=32$, the data are compared to our
finite-size formula; the only fitting parameter is an overall
scale factor.  }
\label{fig:cc_128}
\end{figure}

To test the dipole form, we have carried out Monte Carlo simulations
using the pocket algorithm \cite{mcdimer,dress} on large cubic lattices of
size $L^3$ with $L$ up to 128, with periodic boundary conditions.  We have
computed the connected correlation function for dimers at site $\x$
and at $\x + {\bf r}$, both pointing in the same direction ${\bf
e}_1=[100]$. The vector ${\bf r}$ was taken as a multiple of lattice vectors
$[111]$ $[010]$, $[100]$, $[110]$.  The correlations nicely
fall off as $1/r^\zeta$ in the regime $a\ll r \ll L$, with $\zeta$
close to 3. Furthermore, the ratios of the correlations also agree
very well with the predicted dipole form Eq.~(\ref{e:dipole}),
as shown in Fig.~\ref{fig:cc_128}. 

Going beyond this regime, we can even compute the scaling form of
the correlations for $r \sim L$ through a proper treatment of the
periodic boundary conditions used by our algorithm, which explores
all topological sectors.  In the long wavelength description, we can
write the fields in a given flux (topological) sector as $\bB(\x) =
\sum_i {(\Phi_i/ L^2)} {\bf e}_i +  \bB'(\x)$ so that the field $\bB'$
now carries zero flux. When $ {\bf \Phi} =\sum_i \Phi_i {\bf e}_i$ is
nonzero, the stiffnesses $K_{\parallel}$ and $K_\perp$ for fluctuations
parallel and perpendicular to it will no longer be equal. Nevertheless
these are semi-microscopic and hence can only depend on the average local
magnetic field. Consequently we expect that both $K_{\parallel, \perp}
= K + O(\Phi^2/L^4)$ where $K$ is the stiffness in the zero flux sector.

With these considerations we can generalize (\ref{e:conj}) to
\bea
P[\bB] &\propto& 
e^{ -{K_\parallel \over 2} {\Phi^2 \over L} }
e^{-{1 \over 2} 
\int_V ( K_\parallel B_\parallel^{'2} + K_\perp B_\perp^{'2} )}  \ .
\eea
The first factor will ensure that $\Phi \sim O(\sqrt{L})$ whence
$K_{\parallel, \perp} = K + O(1/L^3)$ so that the stiffness
anisotropy can be ignored for large systems.

We can now deduce the correlations in a given flux sector,
$\langle B_i(\x) B_j(0) \rangle_{\bf \Phi} = {\Phi_i \Phi_j/ L^4}
+ \langle B'_i(\x) B'_j(0) \rangle
$
where the second piece is independent of $\bf \Phi$.  After averaging the
first term with weight $e^{-{K \over 2} {\Phi^2 \over L}}$, it equals
${\delta_{ij}/(K L^3)}$.  From dimensional analysis we know that the
second piece is of the form $f_{ij}(\x/L)/ (K L^3)$ so that the two terms
together have the appropriate form for a finite size scaling function.

Finally, we have also determined the correlator $\langle B'_i(\x) B'_j(0)
\rangle$ in a finite-sized sample subject to periodic boundary conditions.
We are not aware of a closed form evaluation of this quantity
but can write it as a sum in momentum space,
which needs ultraviolet regularization. As we eventually compare it
to a lattice simulation, it is most convenient to do this via a lattice
sum
The resulting forms, e.g.
$\langle B_x({\bf m} a) B_x(0) \rangle={{\cal L}/(K L^3)}$ with ${\cal L}$ 
given by
\bea
1 + \sum_{\bf n\neq{\bf 0}} { [4 - 2 \cos(2 \pi an_y/L) -  2 \cos(2 \pi an_z/L)]
\over [6 - 2 \sum_i \cos(2 \pi an_i/L)]}  e^{i 2 \pi a{{\bf n \cdot m}} \over
L} ~,
\nonumber 
\eea
can then be compared directly with the simulations' results for
distances large compared to a lattice constant, $a$.  In Fig.~\ref{fig:cc_128},
the results are shown for a system of size $L=32$.
After adjusting the one free parameter to $Ka^3=4.9$ the
curves agree well with simulation data for $r$ larger 
than a few lattice spacings. 
A more detailed finite-size analysis that lets
the exponent $\zeta$ vary finds it to be $\zeta=3.00\pm 0.02$,
in excellent agreement with the dipolar form.

The gauge representation can be used to compute other operators of
interest.  Most notably, it shows that the interaction between two
monomers is an attractive entropic force with the same form as that
between oppositely charged monopoles, i.e. an inverse squared force. One
can also consider the Wilson loop 
$\langle\exp({i \int_C \bA \cdot d{\bf l}})\rangle$
along a closed contour C.  In the deconfined phase this will exhibit a
perimeter law at large loop sizes, i.e. it will decay as $e^{-P}$ where
$P$ is the perimeter of the loop C, although with interesting corrections
coming from the long range nature of the Coluomb force \cite{kogutrmp}.

\subsub{FCC Lattice} As was emphasized above, we expect a fundamental
distinction between bipartite and non-bipartite dimer models. For the
latter, the lack of a gauge representation indicates the absence of
a Coulomb phase. To investigate this, we have simulated dimers on the
face-centred cubic lattice, a simple non-bipartite $3d$ Bravais lattice.

\begin{figure}[htbp]
\begin{center}
\epsfig{file=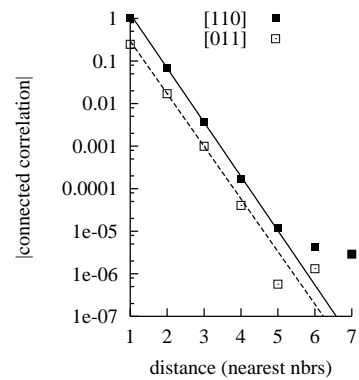,width=0.55\columnwidth}
\end{center}
\caption{ Monte Carlo data for normalized connected dimer correlations,
$|144\langle n_i(\x)n_i(0)\rangle -1|$ on the $L=10$ FCC
lattice, 
for $i$ along the $[110]$ direction.  Separations $\x$ are along $[110]$
and $[011]$ and are measured in nearest-neighbour spacings,
as indicated.}
\label{fig:oozcorr}
\end{figure}

In Fig.~\ref{fig:oozcorr}, we display the connected correlations between
two parallel dimers in two different directions on a system of size 
$L=10$, \emph{i.e.}\ containing $4 \times 10^3$ sites.  The decay of the
oscillating correlations is \emph{extremely} rapid, and fits well to
an exponential form  with a correlation length of $\xi=0.35 \pm 0.01$
nearest neighbour distances. This establishes that the FCC dimer model
is in an exponentially deconfined phase.

\begin{figure}[htbp]
\begin{center}
\epsfig{file=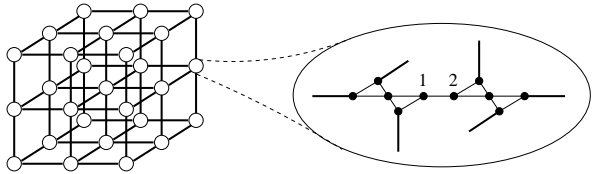,width=0.9\columnwidth}
\end{center}
\caption{ The 3d Fisher lattice is obtained by decorating the sites of
a simple cubic lattice (left) with the cluster shown on the right. 
Its six
external legs (heavy lines) correspond to the original bonds of the
cubic lattice. The numbers denote the possible 
locations of the test monomers.}
\label{fig:fish3d}
\end{figure}

\subsub{Fisher Lattice} It is also possible to identify a $3d$
dimer model with exponential correlations and both confined
and deconfined phases analytically.  This is done by mapping the $3d$
Ising magnet on the cubic lattice onto a dimer model on
the decorated cubic lattice shown in Fig.~\ref{fig:fish3d}, where the
original bonds have fugacity $z=1/\tanh K$ ($K$ being the strength of
the ferromagnetic coupling) whereas those internal to the decorating
clusters have fugacity $1$.  The correlator of the cubic Ising model
between spins $S_i$, $S_j$ at sites $i$ and $j$ can be expressed in terms
of monomer correlators of the resulting dimer model in the same way as
described in detail in Ref.~\onlinecite{fishdim} for the corresponding
$2d$ lattice \cite{fishlat} -- in fact, the analysis carries
over  to any dimension.

One obtains $\left<S_i S_j\right>=\sum_{k_i,k_j=1}^2 m_{(i,k_i),(j,k_j)}$,
where $m_{(i,k_i),(j,k_j)}$ denotes the correlator of test monomers
located on sites $k_i$ of cluster $i$ and $k_j$ of cluster $j$ (see
Fig.~\ref{fig:fish3d}). This correlator is given by the ratio of the
partition functions with and without the pair of monomers present.

In the high temperature phase of the Ising model, $\left<S_i
S_j\right>\rightarrow 0$ exponentially, so that all monomer pairs
are confined. At low temperatures $\left<S_i S_j\right>$ decays
exponentially to a constant. Therefore at least one pair of monomers
is deconfined. Unless possible algebraic terms present in the four
correlators exactly cancel, which seems very unlikely, the deconfinement
is exponential.

\subsub{Conclusions} We have argued that the distinction between the
behaviour of bipartite and non-bipartite lattices familiar from dimer
models in $2d$ holds also in $3d$. The former are characterized by a
conservation law in a long wavelength description which gives us a
Coulomb phase in $3d$ and {\em mutatis mutandis} should do so in all
$d>3$ as well.  Non-bipartite lattices lack such a constraint and can
be generically expected to exhibit exponentially confined or
deconfined phases as they do in our $3d$ examples.  The Coulomb phase
should also arise in other problems that involve conservation laws,
e.g. the statistical mechanics of spin ice which is also the Ising
antiferromagnet on the pyrochlore lattice \cite{henley}. Finally, we
note that Hermele, Fisher and Balents \cite{hermele03} are considering
a set of closely related three-dimensional models.

\subsub{Acknowledgments} RM would like to thank Leon Balents, Matthew
Fisher and Mike Hermele for several useful discussions.  RM was in
part supported by the Minist\`ere de la Recherche et des Nouvelles
Technologies with an ACI grant, and thanks the KITP for hospitality
during part of this work.  The NSF supported this work via 
PHY-9907949 (KITP), DMR-9978074 (SLS) and DMR-0213706 (DAH and SLS). SLS
would also like to acknowledge support by the 
David and Lucile Packard Foundation.

\end{document}